\documentclass[pss]{wiley2sp} 
\usepackage{amsmath}

\begin{document}

\title{Dzyaloshinskii-Moriya interactions and magnetic texture in the Fe
  films deposited on transition-metal dichalcogenides}

\titlerunning{Dzyaloshinskii-Moriya interactions and magnetic texture }

\author{%
  S.~Polesya\textsuperscript{\Ast,\textsf{\bfseries 1}},
  S.~Mankovsky\textsuperscript{\textsf{\bfseries 1}},
  D.~K\"odderitzsch\textsuperscript{\textsf{\bfseries 1}},
W.\ Bensch\textsuperscript{\textsf{\bfseries 2}}, and
  H.~Ebert\textsuperscript{\textsf{\bfseries 1}}
}

\authorrunning{S.~Polesya et al.}

\mail{e-mail
  \textsf{Svitlana.Polesya@cup.uni-muenchen.de}}

\institute{%
  \textsuperscript{1}\, Department Chemie,
  Ludwig-Maximilians-Universit\"at M\"unchen, 81377 M\"unchen, Germany\\
\textsuperscript{2}\, Inst.\ f\"ur Anorgan.\ Chemie,  Universit\"at Kiel,
Olshausenstr.\ 40,  24098 Kiel, Germany
}
\received{XXXX, revised XXXX, accepted XXXX} 
\published{XXXX} 


\abstract{%
  \abstcol{%
    The magnetic properties of materials based on two-dimensional transition-metal
    dichalcogenides (TMDC), namely bulk Fe$_{1/4}$TaS$_2$ compound as
    well as TMDC monolayers with deposited Fe films, have been
    investigated by means of  first-principles DFT 
    calculations.
    Changing the structure and the composition of these two-dimensional
    systems resulted in considerable variations of their physical properties.
}
{
  For the considered systems the Dzyaloshinskii-Moriya (DM) interaction
  has been determined and used for the subsequent investigation of their magnetic
structure using  Monte Carlo simulations. Rather strong DM 
interactions as well as large  $|\vec{D}_{01}|/J_{01}$ ratios have been
obtained in some of these materials, which can lead to the formation of skyrmionic structures varying
with the strength of the applied external magnetic field.
}}

\maketitle   

\section{Introduction}

Magnetic systems with broken inversion symmetry and strong spin-orbit 
interaction exhibit various types of chiral magnetic texture caused by the
interatomic Dzyaloshinskii-Moriya (DM) interaction, namely: helimagnetic structures,
domain walls with a preferred chirality \cite{TRJ+12,HMF+13},
magnetic skyrmions, \emph{etc}. 2D systems like magnetic 
surfaces, interfaces or deposited ultrathin films are considered as
promising candidates for technological applications as they
exhibit the needed properties
required to stabilize  skyrmionic structures, as e.~g. large DM
interactions and strong magnetic anisotropy. In the case of films and interfaces the properties
are essentially determined by the substrate type and structure
parameters, as well as by the type of of magnetic atoms.

Possible candidates for such systems are 3D transition metal dichalcogenide (TMDC)
systems with quasi-2D properties, which are composed of well separated
$TX_2$-'sandwiches' (where $T$ is a transition element atom, and $X$ a
chalcogen atom, S, Se or Te) coupled to each other due to rather weak van der Waals
interactions. Crystallizing in different structure modifications they 
show a rich  variety of interesting physical properties
(mechanical, transport, optical, 2D charge-density wave (CDW) transitions, superconducting
transitions) that are also strongly dependent on chemical composition 
\cite{WY69,WSM75,FY87}.
In recent years a lot of work was devoted to hexagonal TMDC
monolayer systems having outstanding properties   
attractive for spintronic or 2D electronics applications, e.g. the 
formation of direct band gap semiconductor properties of Mo$X_2$ and
W$X_2$, SOC induced valley Hall effect occurring due to the lack of inversion
symmetry \cite{MLH+10,XLF+12,ZKMH14,Hei15}, \emph{etc}.  
The strong spin-orbit coupling (SOC) observed in these systems is
due to the   strong SOC of the heavy metal atoms.
As it was demonstrated theoretically, also in bilayers systems
the properties of the band gap (width, direct/indirect character) can be tuned, if
the bilayers are composed artificially of different TMDC monolayers
\cite{TLT13}.   

It should be noted, that most of the 3D TMDC materials as well as their monolayers
discussed in the literature are non-magnetic. 
However, they 
allow intercalation by a variety of molecules or atoms \cite{FY87} which can
be magnetic, thereby creating a class of magnetic materials based on TMDC
\cite{PF80,PF80a,EMD+81,NIH+94}. 
In the case of TMDC surfaces or films two-dimensional magnetic systems can be created
by deposition of magnetic overlayers. Because of lack of inversion
symmetry in such systems one can expect strong DM interactions which can
be tuned by different elements in the TMCD substrate.
Even though no experimental measurements have been performed so far for these
systems, it is worth to investigate their properties, in particular,
their potential for skyrmion formation, to be subsequently used as a
material for memory storage devices.

\section{Computational details}

Within the present work, spin-polarized  electronic structure
calculations for the ground-state have  been performed using the
multiple scattering KKR (Korringa-Kohn-Rostoker) Green function method
~\cite{SPR-KKR6.3,EKM11} in the scalar-relativistic
approximation. The local spin density approximation (LSDA) to the exchange-correlation functional in
 density functional theory was used with the 
parametrisation of the exchange-correlation potential by Vosko,
Wilk, and Nusair  \cite{VWN80}. 
The potential geometry was treated within the
atomic sphere approximation (ASA). For the angular momentum
expansion of the Green function a cutoff of
$l_{max} = 3$ was applied.  
The isotropic exchange coupling parameters and components of the
Dzyaloshinskii-Moriya interactions have been calculated within the Green
function formalism \cite{EM09a}. These parameters have been used for
subsequent Monte Carlo (MC) simulations based on the extended Heisenberg model
\cite{USPW03}, using the Metropolis algorithm.
The structural parameters for bulk Fe intercalated Fe$_{1/4}$TaS$_2$ was
taken from experiment \cite{MZL+07}, while for the TMDC monolayers with Fe
overlayers the structure parameters ($a$ and $c$) have been obtained within the
structure optimization procedure using the VASP code \cite{KF96}.
More details about Fe intercalated TaS$_2$ systems one can see in Ref. \cite{MCK+15}.

\section{Results}

\begin{figure}
  \begin{center}
\includegraphics[width=0.3\textwidth,angle=0,clip]{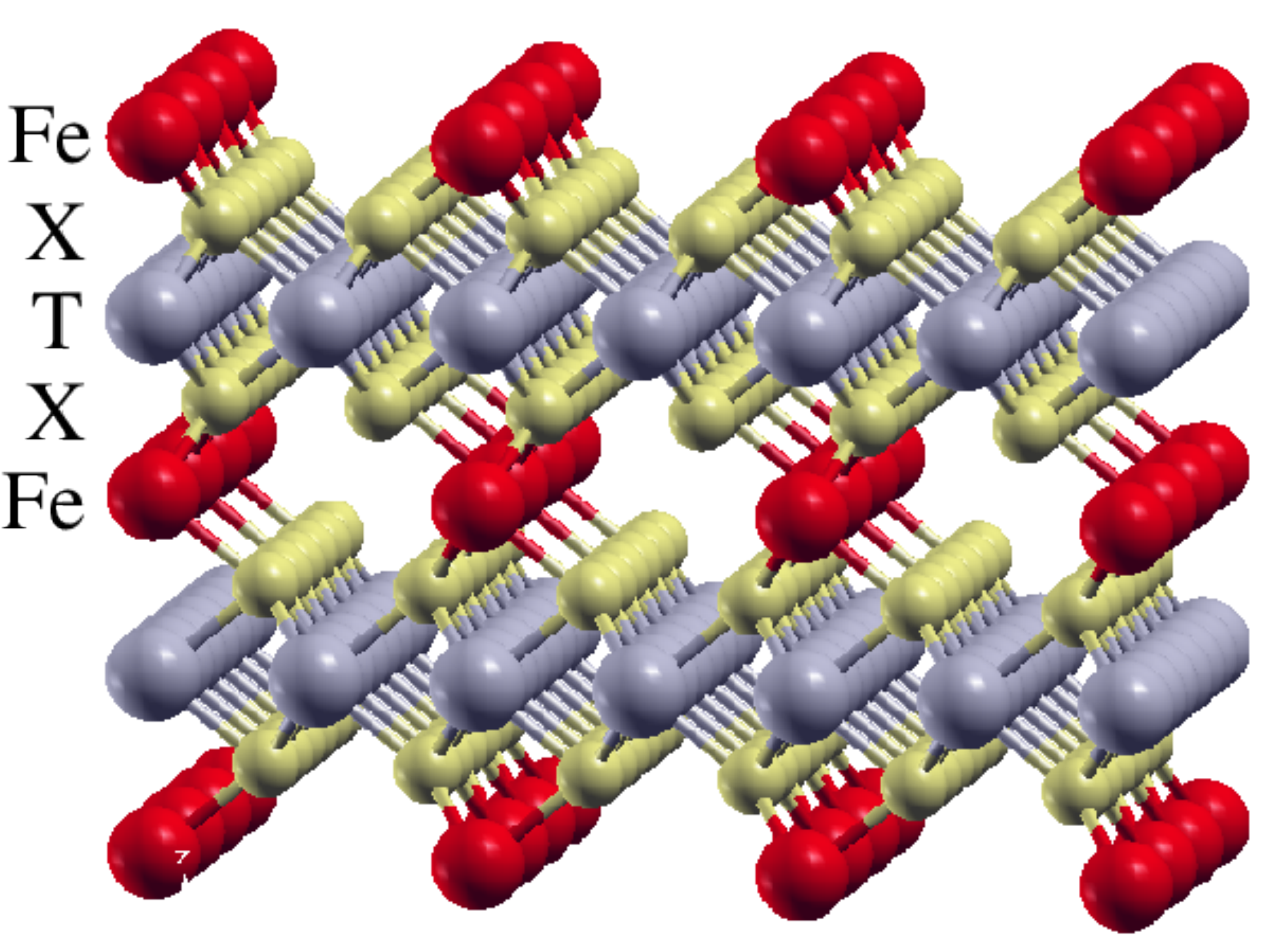}\;(a) 
\includegraphics[width=0.2\textwidth,angle=0,clip]{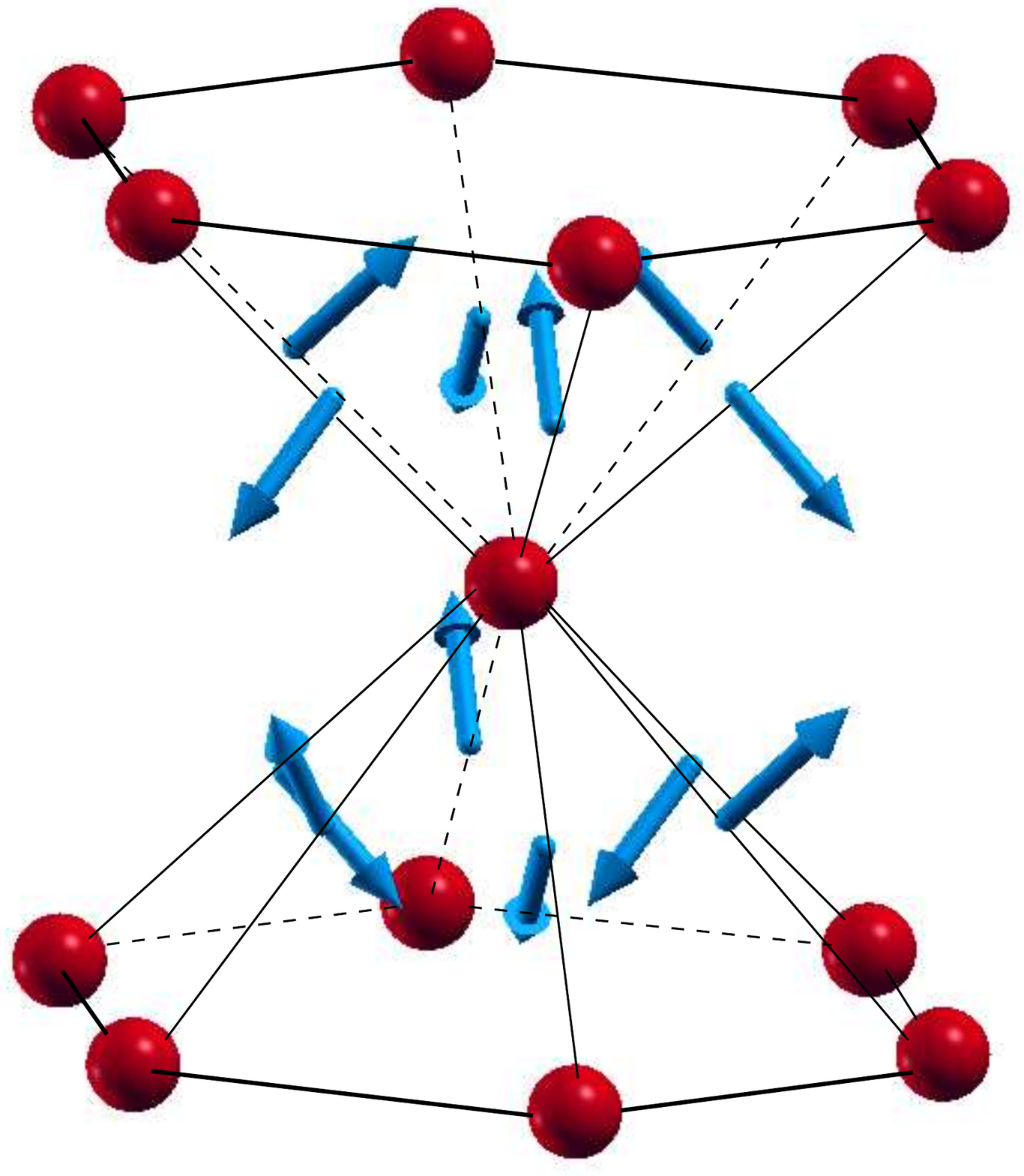}\;(b)
\includegraphics[width=0.2\textwidth,angle=0,clip]{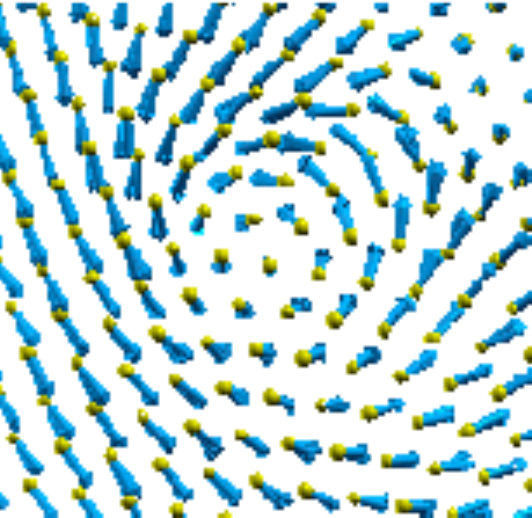}\;(c)
\caption{\label{fig:skyrm}  Fe intercalated 3D TMDC, Fe$_{1/4}$TaS$_2$:
spatial structure (a);
directions of the DM vectors between atom pairs (b); magnetic texture (top view)
obtained in MC simulations for $T=1$\ K in the presence of an in-plane
magnetic anisotropy ($0.1$\ meV) and an 
out-of-plane external magnetic field
$|\vec{B}_{ext}| = 0.2 $\ T (c).
  }  
  \end{center}
\end{figure}

As was pointed out, the TMDC systems exhibit strong SOC effects.
Therefore strong DM interactions in the materials based on TMDC 
without inversion symmetry can be 
expected, that are formed by  intercalation or deposition of
magnetic layers. 
Calculations for the Fe intercalated bulk Fe$_{1/4}$TaS$_2$ (see the
structure in Fig. \ref{fig:skyrm}(c)) have been
performed using the experimental structure parameters. 
Due to modifications of the electronic structure caused by the Fe
intercalation the CDW instability observed for pure TaS$_2$ is removed.
The Fe  spin magnetic moments obtained in our calculations for
Fe$_{1/4}$TaS$_2$ is  $m_{spin} = 2.76~\mu_B$.
Fig. \ref{fig:skyrm}(b)
shows the directions of the first non-zero DM interaction,
$|\vec{D}_{01}| = 0.3$~ meV, between the Fe atoms which belong to
different neighboring Fe layers. Within the 
layers the DM interaction are zero for symmetry reason.
The strongest exchange interaction $J_{01} = 8.0$\ meV is observed for  Fe
atoms which belong to neighboring Fe planes, that, in turn, results in the ferromagnetic (FM)  
alignment of their magnetic moments. Within the planes the isotropic
exchange interactions are rather small, $J_{02} = 0.74$ meV. 
Although
the DM interactions between the atoms arranged within the plane are
zero, one can  expect that the rather strong interlayer DM
interactions leads to the  formation of a pronounced magnetic texture.

MC simulations have been performed
taking into account  the observed experimentally out-of-plane magnetic
anisotropy in the system.
No skyrmion structure has been obtained at any
magnitude of magnetic field. Note, however, that an artificial rotation of
the direction of the magnetocrystalline anisotropy (MCA) into the plane results in the formation of a
half-skyrmion (or meron) magnetic texture. This can be seen in the
snapshot of magnetic configuration (see Fig. \ref{fig:skyrm}(c)) obtained
at 1.0 K with the MCA energy $0.1$\ meV and external magnetic field $|\vec{B}_{ext}| =
0.2$\ T.

\begin{figure}
\includegraphics[width=0.15\textwidth,angle=0,clip]{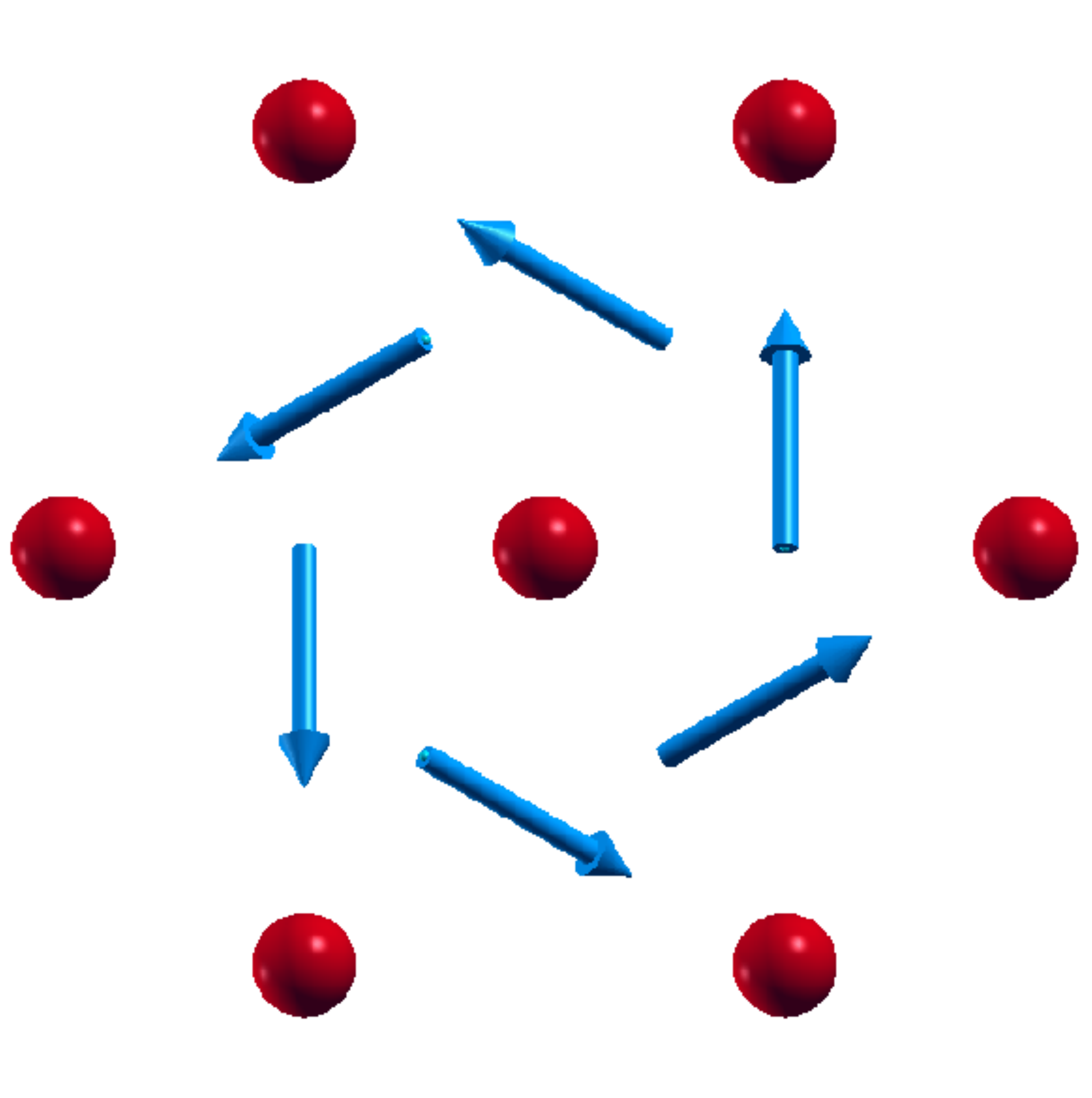}\;(a)
\includegraphics[width=0.3\textwidth,angle=0,clip]{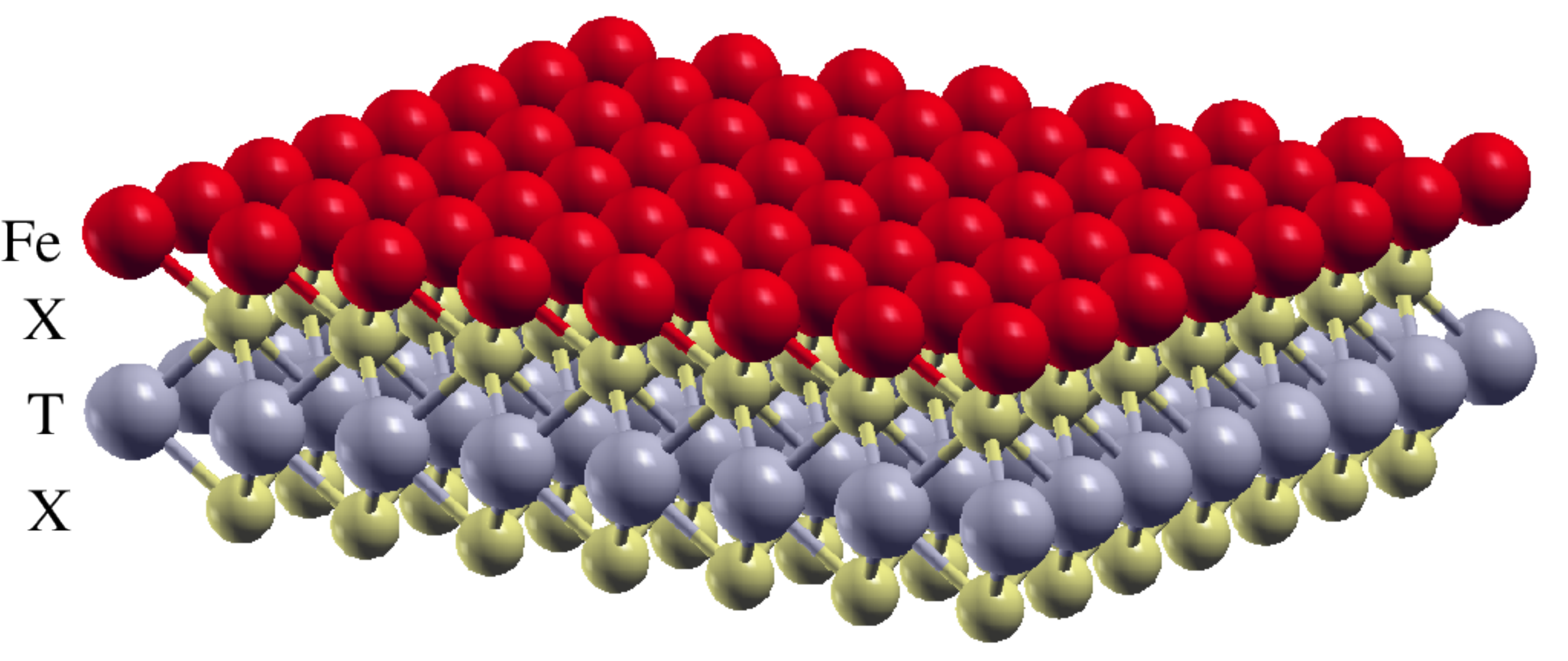}\;(b)\\
\caption{\label{fig:Geom_2D} (a) The directions of DM interactions for
   Fe overlayers deposited on one TMDC monolayer (top view). (b)
   Geometry of the monolayer Fe/1H-TMDC system. In the case of $2 \times
   2$ Fe overlayer, every second position within the layer is occupied
   by Fe atoms. 
 }    
\end{figure}

\begin{figure}
\begin{center}
\includegraphics[width=0.17\textwidth,angle=0,clip]{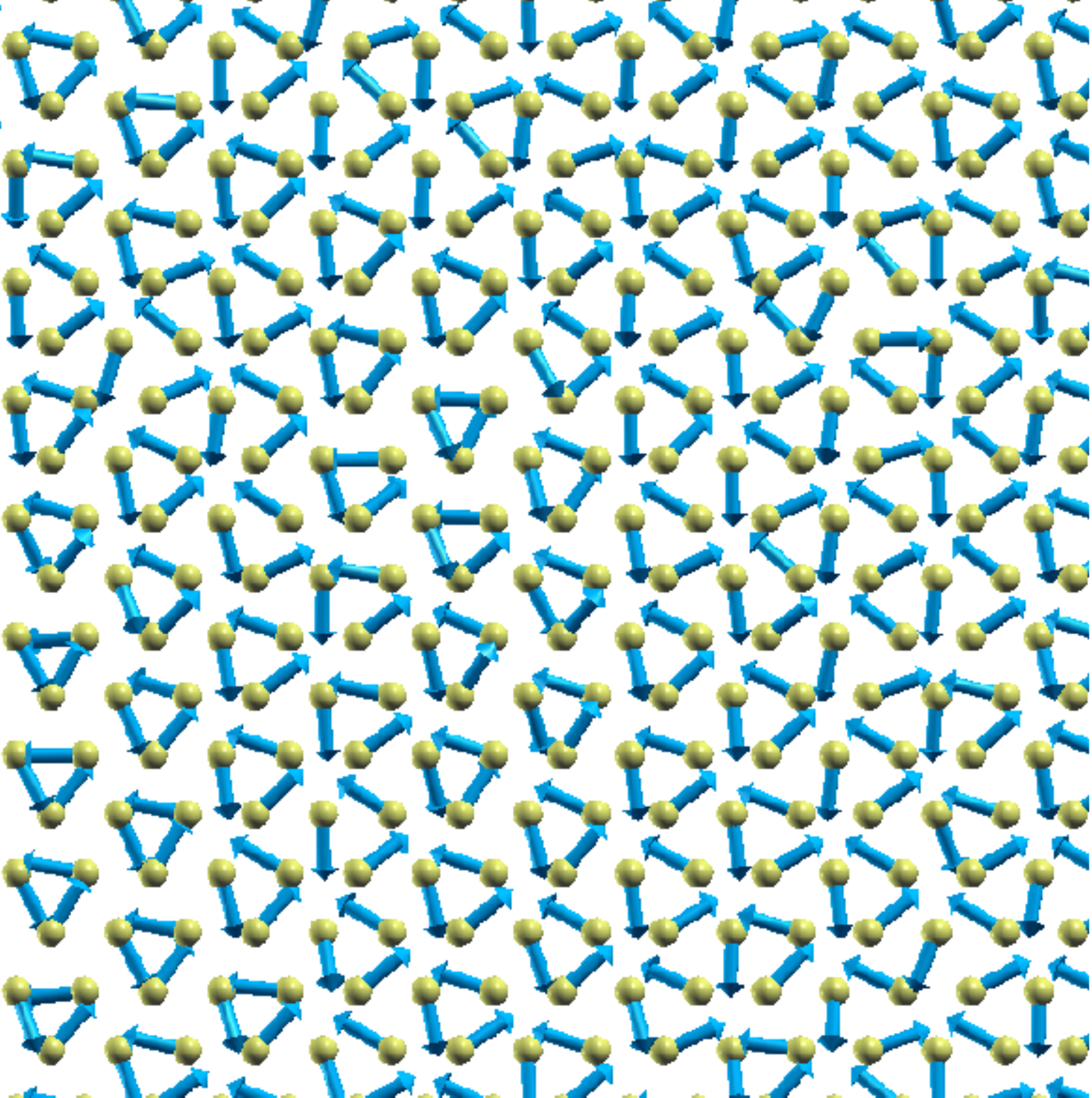}\;(a)
\includegraphics[width=0.17\textwidth,angle=0,clip]{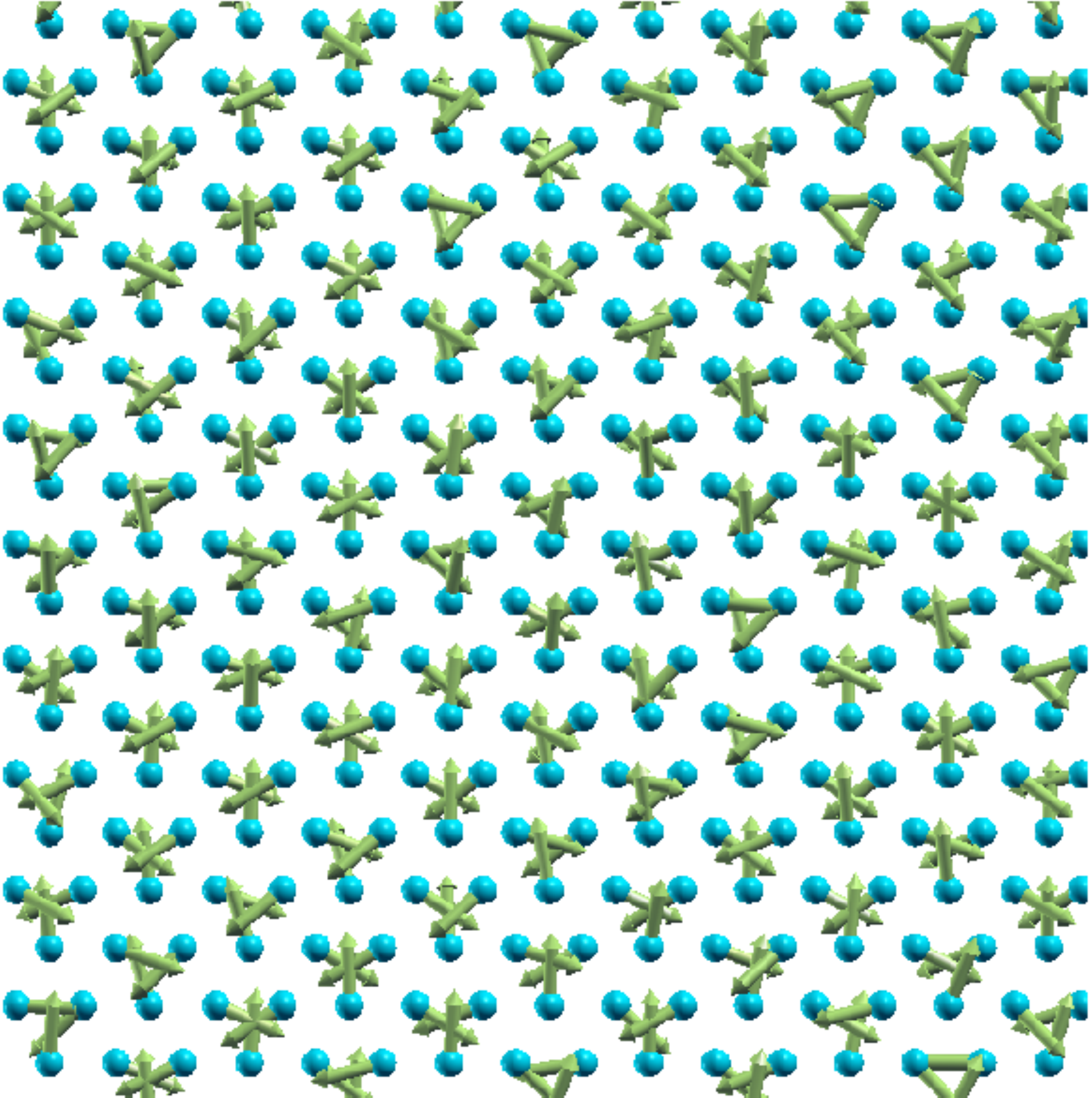}\;(b)\\
\includegraphics[width=0.17\textwidth,angle=0,clip]{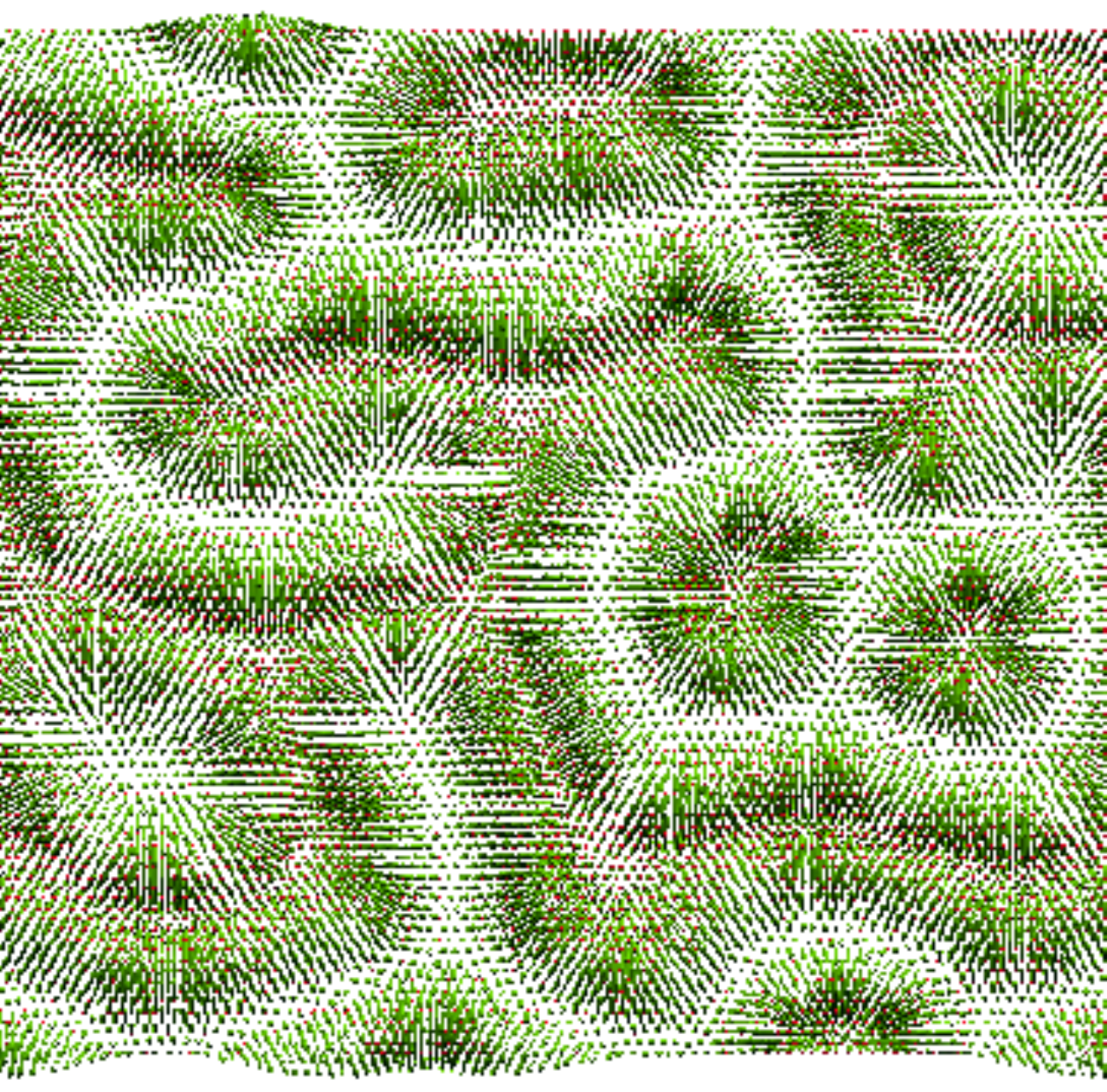}(c)
\includegraphics[width=0.17\textwidth,angle=0,clip]{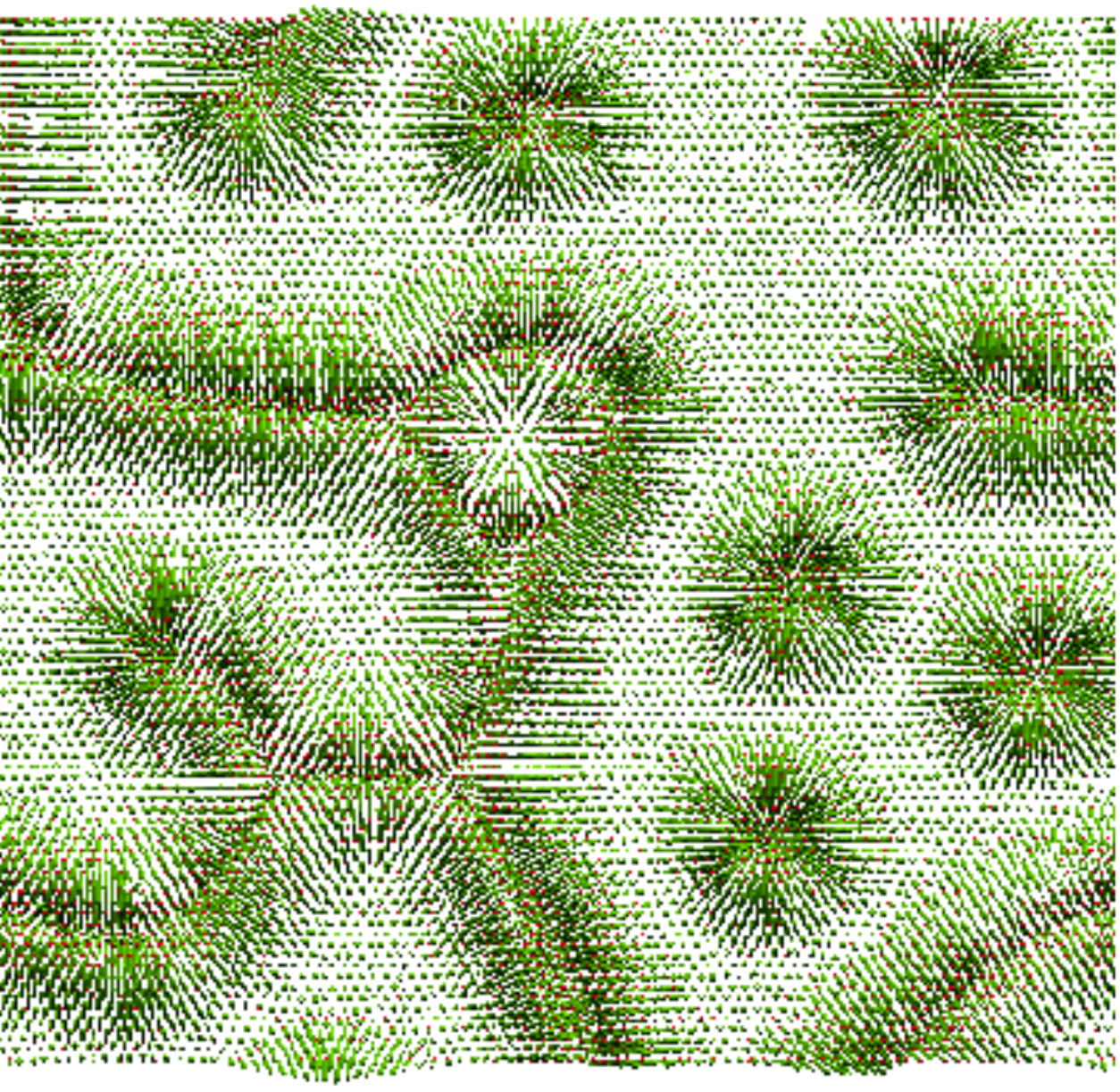}(d)\\ 
\includegraphics[width=0.17\textwidth,angle=0,clip]{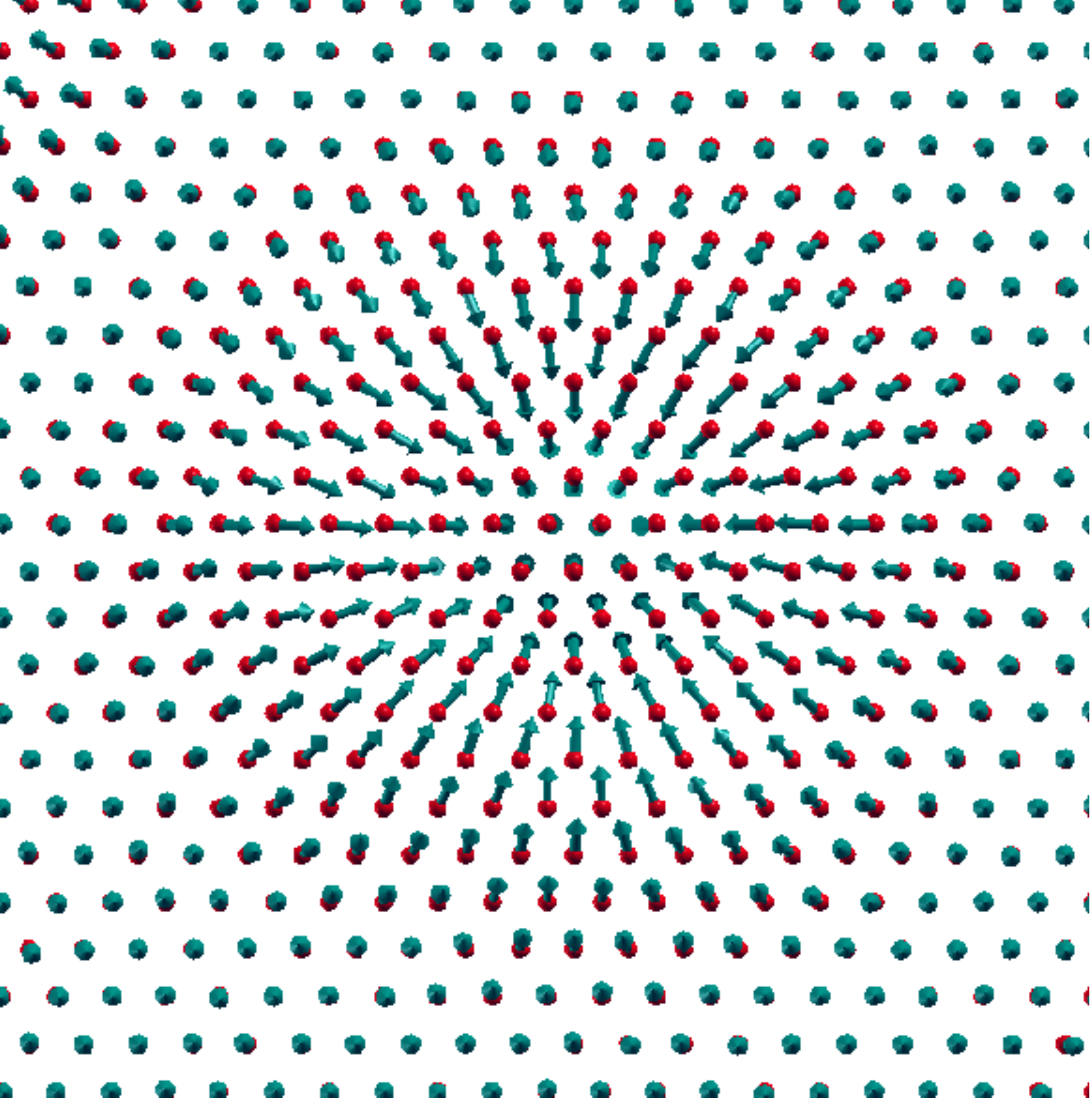}(e)
\includegraphics[width=0.17\textwidth,angle=0,clip]{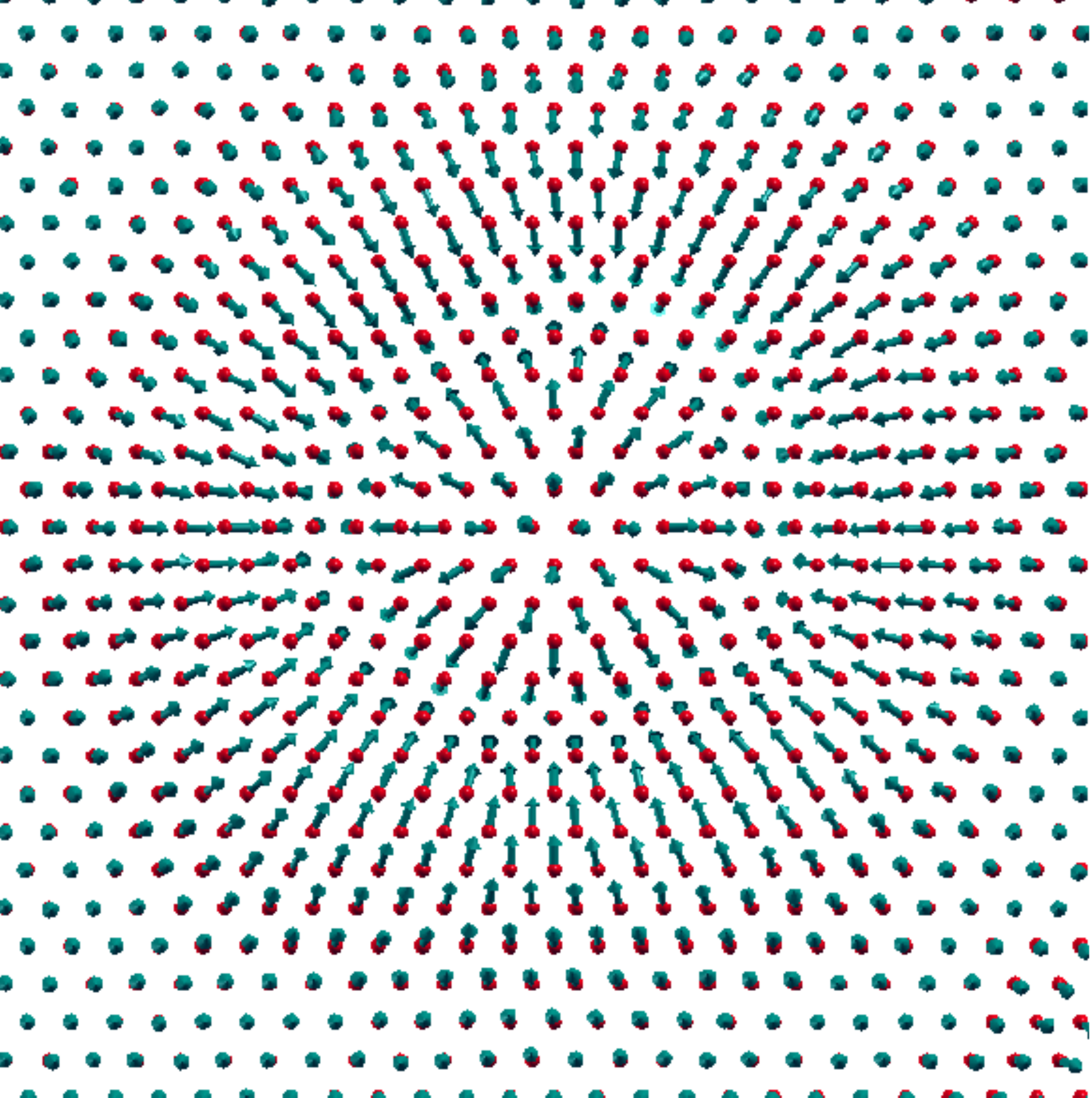}(f)
\caption{\label{fig:Fe_TMDC} 
Snapshot of the magnetic configurations obtained within the MC simulations
for $2 \times 2$ Fe overlayer on the 1H-TMDC monolayer for
$|\vec{B}_{ext}| = 0$~ T (a) and  $0.9$~ T (b).
 The  field-induced modification of the
  magnetic texture in the 1T-based  Fe/TaTe$_2$ system   obtained within
  MC simulations at $T = 0.5$\ K:  $|\vec{B}_{ext}| = 0.0$\ T (c) and
  $2.6$\ T (d). Snapshot of single magnetic skyrmions with winding
  numbers $w=1$ (e) and  $w=2$ (f), obtained within MC simulations for
  Fe/1T-TaTe$_2$. }  
\end{center}
\end{figure}

As mentioned above, one can expect strong DM interactions for
2D systems consisting of Fe overlayers on top of a TMDC monolayer. Therefore,
calculations have been performed for a TaS$_2$ 
monolayer with a deposited Fe overlayer. Magnetic anisotropy 
was not taken into account in MC simulations in the present work.
In a first step, the calculations have been
perfomed for an occupancy of 25\% of the sites within the Fe-layer, assuming that the Fe atoms create
an ordered structure similar to that observed in the intercalated
systems Fe$_{1/4}$TaS$_2$. 
Because of the large distance between the Fe atoms within  the $2 \times 2$
monolayer, the isotropic interactions between the first neighboring Fe 
atoms within the layer are rather small,  $J_{01} = -1.9$\ meV, and
comparable to the DM interactions, $|\vec{D}_{01}| = 0.9$\ meV.
However, the sign of $J_{01}$ is negative, leading to a frustrated
magnetic structure within the Fe layer with the Fe magnetic moments
lying in the film plane. 
The $D_z$ component of the DM interaction vector is nearly zero and  the DM
vectors are oriented almost within the plane as it is shown in Fig.\
\ref{fig:Geom_2D}(a). As a result, at low temperatures the effect of  the DM
interaction is hardly seen Fig.\ \ref{fig:Fe_TMDC}(a). However, when the
temperature increases leading to spin moment fluctuations with
increasing $z$-component of the Fe spin moments, the modulation of the
magnetic structure due to the DM interactions becomes more pronounced.  
An external magnetic field $\vec{B}_{ext}$ along the $z$ direction
(perpendicular to the film plane) forces the magnetic moments to be
aligned parallel to the field, that, in turn, leads to a pronounced modification
of the magnetic structure due to the DM interactions as is shown in
\ref{fig:Fe_TMDC}(b).

The calculations on a deposited Fe monolayer have been performed for TMDC
monolayers with 1T and 1H polytypes with the $T$ atoms occupying the positions
with octahedral and trigonal-prismatic coordination, respectively (see
geometry for Fe/1H-TMDC in Fig. \ref{fig:Geom_2D}(b)).
Some results of the calculations for the systems under considerations are
presented in Table \ref{tab1}. 
As we focus here on the interatomic exchange tensor, one can see
that the isotropic exchange interactions $J = J_{01}$ are larger while
DM $D = |\vec{D}_{01}|$ interactions are smaller for the Fe films deposited on
1H-TMDC monolayers when compared to the Fe/1T-TMDC system. As a result,
the ratio $D/J$ is significantly smaller in the first case. 
As it follows from the calculations for Fe/1T-TMDC system, the transition 
from $X =$~S to Te results in a decrease of $J$ values leading as a
consequence to an increase of the $D/J$ ratio. 
Note that in the case of Fe/1T-WTe$_2$ system, $J_{01}$, $J_{02}$ and
$J_{03}$ are negative and $|J_{03}|>|J_{01}|$. Therefore, the values of
exchange interactions are presented in Table \ref{tab1} for two 
distances, while $D/J$ ratio is omitted as they do not have significant
meaning in this case. The directions of DM vectors in the systems under
consideration are shown in Fig. \ref{fig:Geom_2D}(a).  

Figure \ref{fig:Fe_TMDC}(c)-(d) displays the results of MC simulations for
Fe/1T-TaTe$_2$ obtained at $T = 0.5$\ K and the external magnetic field 
$0$\ T and $2.6$\ T. Without field, one can see a rather complicated
helimagnetic structure (similar to one obtained in \cite{SPR+14}) in
contrast to that observed for example in the case of $(2\times1)$ FePt
monolayers on the Pt(111) surface \cite{PMB+14}. 
An increase of $B_{ext}$ results in an increase of the area of
FM-ordered regions with the magnetization along
$\vec{B}_{ext}$ and a stabilization of magnetic skyrmions (having
winding number one, Fig. \ref{fig:Fe_TMDC}(e)).
In addition, one can see in Fig. \ref{fig:Fe_TMDC}(d) a magnetic texture
with a complicated topology and with the core magnetization along the
magnetic field. This texture can transform at the increasing magnetic field
into the skyrmion with winding number equal to two, $w = 2$ (see
Fig. \ref{fig:Fe_TMDC}(f)). The following increase of $\vec{B}_{ext}$,
however, results in the vanishing of these types of skyrmions leading to
the conventional skyrmionic structure with the skyrmion density varying
with the strength of the magnetic field.

\begin{table}[]
  \begin{center}
 \begin{tabular}{|l|l|l|l|l|c|}
  \hline
           & $R_{WS}$   & $m_s$  & $J$    &  $D_{z} | D_{||}$   & $D/J$  \\
  \hline
 \multicolumn{6}{c}{1H-polytype} \\
  \hline
TaS$_2$   &  2.81  & 2.12  & 13.71 &  $0.73 | 0.63$ & 0.07\\
TaSe$_2$  &  2.94  & 2.15  & 10.76 &  $0.35 | 0.62$ & 0.06\\
WS$_2$    &  2.74  & 1.98  & 19.31 &  $0.41 | 0.49$ & 0.03\\
WSe$_2$   &  2.89  & 1.85  & 16.93 &  $0.72 | 0.85$ & 0.07\\
  \hline
 \multicolumn{6}{c}{1T-polytype} \\
  \hline
 TaS$_2$    & 2.92 & 2.54 & 13.46 & $0.61 | 1.05$  & 0.09 \\
 TaSe$_2$   & 3.07 & 2.57 & 10.79 & $0.82 | 1.73 $  & 0.18 \\
 TaTe$_2$   & 3.28 & 2.45 & 6.90 & $0.33 | 1.77$  & 0.26 \\
  WS$_2$     & 2.86 & 2.57 & 7.22 & $0.60 | 0.31$  & 0.09 \\
  WSe$_2$    & 3.01 & 2.67 & 7.12 &  $0.62 | 1.05$ & 0.17 \\
  WTe$_2$:   &  &  &  &   &  \\
  ($1^{st}$ shell)   & 3.22 & 2.51 & -0.33 & $ 0.84 | 0.85$  & - \\
  ($3^{rd}$ shell)   & 3.22 & 2.51 & -4.37 & $ 1.95 | 0.45$  & - \\
  \hline
  \end{tabular}
  \end{center}
 \caption{The Wigner-Seitz radius, $R_{WS}$ (a.u.) and spin magnetic
   moment $m_s$ ($\mu_B$) of the Fe atom for the TMDC
   monolayers with different structure and composition. $J=J_{01}$ (meV)
   and $D = |\vec{D}_{01}|$ (meV) represents the isotropic and DM exchange
   interactions, respectively.  $D_{||}$ (meV) and $D_{z}$ (meV) are the
   in-plane and out-of-plane components of the DM vector $\vec{D}_{01}$.
   In the case of WTe$_2$ the interactions are presented for the $1^{st}-$
   and $3^{rd}-$neighbor shells.
 }
 \label{tab1}

\end{table}

\section{Summary}

We have demonstrated that rather strong DM interactions as well as
large  $D/J$ ratios can be obtained in  systems based on
TMCD compounds, namely Fe-intercalated bulk or monolayers with
deposited Fe film. 
These systems  are attractive candidates for  spintronic applications
showing large tunablity of their physical properties. 
This was demonstrated by showing  the possibility of the 
formation of skyrmionic structure in these materials. 
A variety of properties have been obtained by changing the structure and
the composition of the systems with Fe overlayers on  top of TMDC
monolayer. Using  Fe/1T-TaTe$_2$ film as an example with a large
$D/J$ ratio, we have demonstrated the 
formation of skyrmions
with a winding number two in these systems observed at certain value of the magnetic
field, which disappear when the strength of magnetic field increases. 
As an outlook, the TMDC monolayers with  antiferromagnetic oriented $3d$-metal films 
on both sides of TMDC can be considered as a prototype of  bi-layered
systems to observe antiferromagnetic skyrmions \cite{ZZE15a}.

\section{Acknowledgements}

Financial support by the DFG via SFB 689 (Spinph\"anomene in reduzierten
Dimensionen) is thankfully acknowledged.


\providecommand{\WileyBibTextsc}{}
\let\textsc\WileyBibTextsc
\providecommand{\othercit}{}
\providecommand{\jr}[1]{#1}
\providecommand{\etal}{~et~al.}

\end{document}